\documentclass[a4paper,superscriptaddress,twocolumn,nofootinbib,longbibliography]{aastex631}
\usepackage{aas_macros}
\usepackage{acro}
\acsetup{patch/longtable=false}
\usepackage{graphicx}
\usepackage{dcolumn}
\usepackage{bm}
\usepackage{comment}
\usepackage{amsmath}
\usepackage{float}
\usepackage{lipsum}
\usepackage{color}
\usepackage{subfigure}
\usepackage{times}
\usepackage{textcomp}
\usepackage{tabularx}
\usepackage{latexsym}
\usepackage{mathtools}
\usepackage{url}
\usepackage[utf8]{inputenc}
\usepackage[normalem]{ulem}
\usepackage{physics}
\usepackage{subfigure}
\usepackage{soul}
\usepackage{hyperref}
\hypersetup{colorlinks,linkcolor={blue},citecolor={blue},urlcolor={red}}  
\setcitestyle{authoryear,open={},close={}}
\setcitestyle{open={(},close={)}}
\usepackage{color}
\definecolor{mygreen}{rgb}{0.1,0.8,0.1}

\begin{document}

\title{Residual test to search for microlensing signatures in strongly lensed gravitational wave signals.}

\author{Eungwang Seo}
\thanks{e.seo.1@research.gla.ac.uk}
\affiliation{SUPA, School of Physics and Astronomy, University of Glasgow, Glasgow G12 8QQ, United Kingdom}

\author{Xikai Shan}
\affiliation{Department of Astronomy, Tsinghua University, Beijing 100084, China}

\author{Justin Janquart}
\affiliation{Center for Cosmology, Particle Physics and Phenomenology - CP3, Universit\'e Catholique de Louvain, Louvain-La-Neuve, B-1348, Belgium}
\affiliation{Royal Observatory of Belgium, Avenue Circulaire, 3, 1180 Uccle, Belgium}

\author{Otto Hannuksela}
\affiliation{Department of Physics, The Chinese University of Hong Kong, Shatin, NT, Hong Kong}

\author{Martin Hendry}
\affiliation{SUPA, School of Physics and Astronomy, University of Glasgow, Glasgow G12 8QQ, United Kingdom}

\author{Bin Hu}
\affiliation{School of Physics and Astronomy, Beijing Normal University, Beijing 100875, China}

\begin{abstract}
When a gravitational wave signal encounters a massive object, such as a galaxy or galaxy cluster, it undergoes strong gravitational lensing, producing multiple copies of the original signal.
These strongly lensed signals exhibit identical waveform morphology in the frequency domain, allowing analysis without the need for complex lens models. 
However, stellar fields and dark matter substructures within the galactic lens introduce microlensing effects that alter individual signal morphologies.
Identifying these microlensing signatures is computationally challenging within Bayesian frameworks.
In this study, we propose a residual test to efficiently search for microlensing signatures by leveraging the fact that current Bayesian inference pipelines are optimized solely for the strong lensing hypothesis. 
Using cross-correlation techniques, we investigate the microlensing-induced deviations from the strong hypothesis, which are imprinted in the residuals.
Most simulated signals from our realistic microlensing populations exhibit small mismatches between the microlensed and unlensed waveforms, but a fraction show significant deviations.
We find that 28\% (52\%) and 34\% (66\%) of microlensed events with mismatch $\geq 0.03$ and $\geq 0.1$, respectively, can be discerned with O4 (O5) detector sensitivities, which demonstrates that high-mismatch events are more likely to be identified as microlensed.
Including all events from a realistic population, 11\% (21.5\%) are identifiable with O4 (O5) sensitivity using our approach.
\end{abstract}

\section{Introduction} \label{sec:intro}
Like electromagnetic waves, gravitational waves (GWs) experience gravitational lensing when they propagate near massive objects, leading to waveform amplifications, distortions, phase shifts, and changes in their propagation paths, depending on the lens-source system characteristics and the GW wavelength~\citep{ohanian1974focusing,Thorne:1982cv,deguchi1986diffraction,wang1996gravitational,nakamura1998gravitational,takahashi2003wave}.

GW lensing can be broadly classified into three main categories based on the strength and scale of the lensing effect: strong lensing, weak lensing, and microlensing.
Strong lensing involves large-scale objects, such as galaxies or galaxy clusters, which produce multiple copies of the original GW signal---referred to as \emph{images}~\citep{PhysRevD.97.023012,li2018gravitational,oguri2018effect,smith2017strong,smith2018if,smith2019deep,robertson2020does,ryczanowski2020building}.
Strongly lensed signals undergo (de-)magnifications, time delays, and overall phase shifts that do not affect the frequency evolution~\citep{wang1996gravitational,dai2017waveforms,PhysRevD.103.064047}.
Similarly, weak lensing also arises from the intervening large-scale distribution of matter, but the resulting deflections are subtle and do not produce multiple images, leading only to an overall (de)magnification~\citep{mukherjee2020probing,mukherjee2020multimessenger}.
Microlensing, in contrast, is caused by smaller-scale masses, such as individual stars or compact objects, inducing frequency-dependent wave-optics effects on the waveforms~\citep{deguchi1986diffraction,nakamura1998gravitational,takahashi2003wave,PhysRevD.90.062003,PhysRevD.98.083005,PhysRevD.98.103022,PhysRevLett.122.041103}.

Approximately $\mathcal{O}(1)$ strongly lensed GW signals are expected to be detected with the LIGO-Virgo-KAGRA (LVK) detectors at their design sensitivities, while third-generation detectors are anticipated to observe around $\mathcal{O}(10)$ such events~\citep{ng2018precise,li2018gravitational,wierda2021beyond}.
The LVK Collaboration performed GW strong lensing and microlensing searches for binary black holes (BBHs) detected during the O1 to O3 observing runs, not finding convincing evidence for lensing so far~\citep{hannuksela2019search,abbott2021search,abbott2023search,janquart2023follow}.
In these searches, the strong lensing pipelines are typically model-independent and rely on geometrical optics~\citep{Haris:2018vmn,mcisaac2020search,liu2021identifying,goyal2021rapid,janquart2021fast,lo2023bayesian,ezquiaga2023identifying,janquart2023return,goyal2024rapid,magare2024slick,chakraborty2024glance,gholap2025chi,barsode2025fast} and the microlensing pipeline assumes the presence of a single axially-symmetric microlens~\citep{wright2022gravelamps}.

In a more realistic scenario, a GW signal propagating through a galaxy or a galaxy cluster is likely to undergo both strong lensing from the main lensing object and microlensing from the stellar fields or dark matter substructures embedded within it, which may leave more complex imprints on the GW images~\citep{diego2019observational,PhysRevD.101.123512,cheung2021stellar,mishra2021gravitational,meena2022gravitational,seo2022improving,yeung2023detectability,shan2023microlensing,Shan:2023qvd}.
If the microlensing effects are not accounted for in lensing search pipelines, lensed GW signals may go undetected~\citep{mishra2024exploring,chan2024detectability} 
However, incorporating complicated lensing effects into existing pipelines could significantly increase computation time, making them inefficient for analyzing the numerous GW events expected to be observed by future detectors~\citep{baibhav2019gravitational,Samajdar:2021egv,iacovelli2022forecasting}.

To bypass this problem, we propose conducting ``residual tests'' instead of traditional Bayesian inference methods to identify evidence for microlensing in strongly lensed GW signals.
Specifically, we investigate the residuals left in the observed data after subtracting the best-fit waveform template---similar to the approach used in studies testing general relativity (GR), which aim to identify deviations from GR-modeled templates in observed signals~\citep{abbott2016tests,abbott2019tests,abbott2021tests2,abbott2021tests3}.

In the following sections, we introduce the lensing framework and how to conduct residual tests.
In Sec.~\ref{sec:analysis}, we introduce how microlensed GW signals are described, after which we
detail the methodology of our residual test in Sec.~\ref{sec:residualtest}.
We present the results of our studies in Sec.~\ref{sec:result}, and discuss the implications of our findings for future microlensing analyses in Sec.~\ref{sec:discussion}.

\section{Analysis on lensed gravitational waves}\label{sec:analysis}

\subsection{Gravitational wave lensing}\label{framework}
\begin{figure}[t]
    \centering
\includegraphics[width=1.0\linewidth]{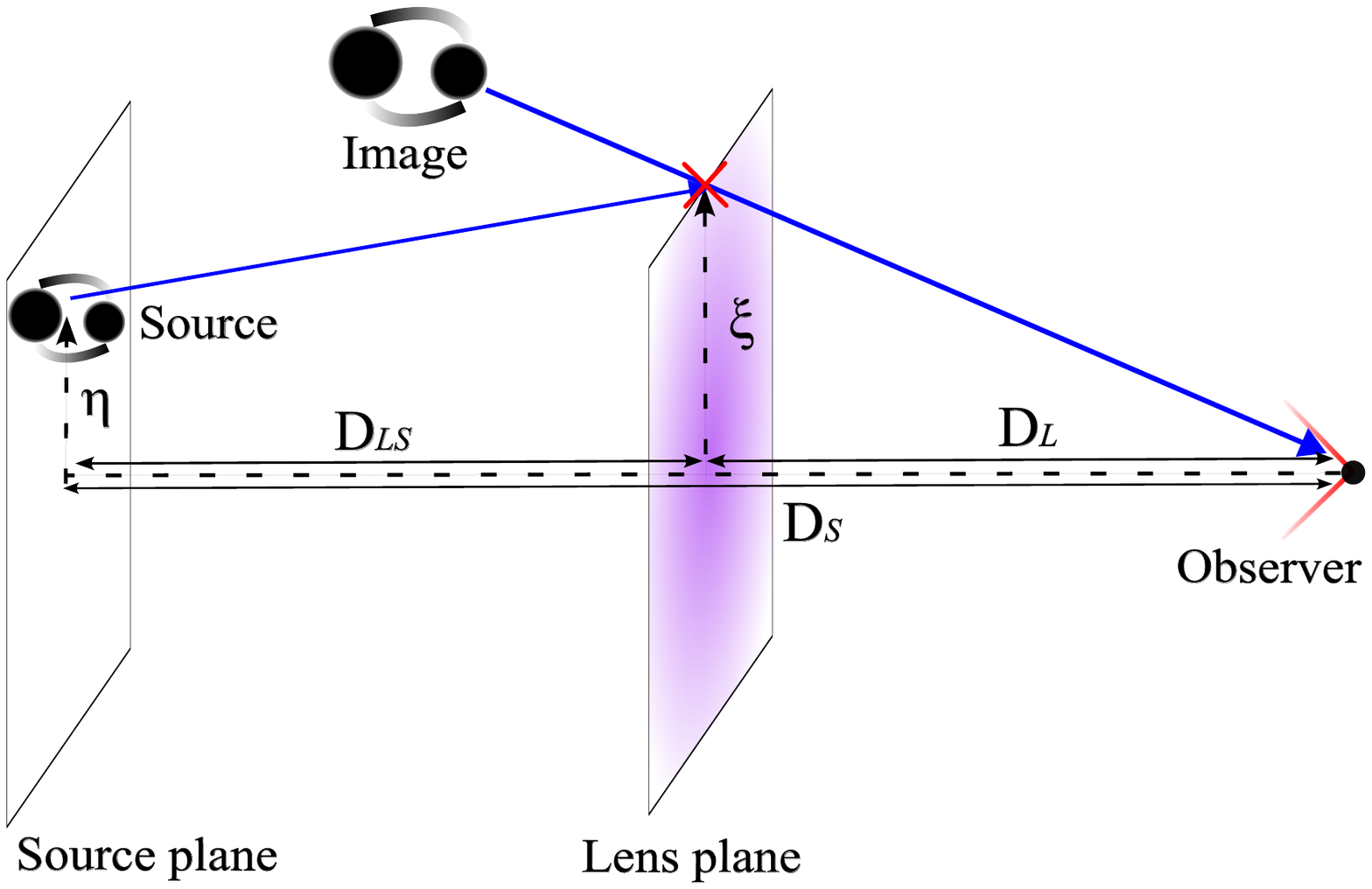}
    \caption{Schematic diagram of GW lensing geometry.
    A GW emitted from a distant source is deflected by a foreground lens before arriving at the observer.
    $\boldsymbol{\xi}$ denotes the deflected image position from the lens center (red cross) in the lens plane, while $\boldsymbol{\eta}$ represents the source position from the optical axis in the source plane.
    $D_{L}$, $D_{S}$, and $D_{LS}$ are the angular diameter distances between observer and lens, observer and source, and source and lens, respectively.
    }
    \label{lensing_geometry}
\end{figure}

In this work, we adopt the thin lens approximation, in which the thickness of the lens is negligible compared to the distances between the source, lens, and observer.
Under this assumption, all lensing effects are confined to a single plane, referred to as the lens plane.
The geometry of a GW lensing system is illustrated in Figure~\ref{lensing_geometry}.

For convenience, the image and source positions are expressed in dimensionless form.
\begin{equation}
    \boldsymbol{x}=\frac{\boldsymbol{\xi}}{\xi_{0}}, \:\:
    \boldsymbol{y}=\frac{D_{L}\boldsymbol{\eta}}{D_{S}\xi_{0}},
\label{position}
\end{equation}
where $\boldsymbol{\xi}$ and $\boldsymbol{\eta}$ denote physical displacements from the lens center to an image in the lens plane, and from the optical axis to the source in the source plane, respectively.
$\xi_{0}$ is a normalization constant used to render the dimensionless coordinates.
$D_{L}$ and $D_{S}$ are the angular diameter distances between the observer and the lens, and the observer and the source, respectively.

The waveforms of lensed GW signals, $h_{L}(f)$, can be calculated by multiplying the unlensed signal, $h_{U}(f)$, by a frequency-dependent amplification factor $F(f)$.
Regardless of which lens model is assumed, the $F(f)$ can be obtained by calculating the Fresnel-Kirchhoff diffraction integral as follows,~\citep{deguchi1986diffraction,nakamura1998gravitational,takahashi2003wave} 
\begin{equation}
\label{af}
F(f, \boldsymbol{y})=\frac{4 G M^{z}_L f}{c^3 i} \int_{-\infty}^{\infty} \mathrm{d}^2 x \exp [2\pi i ft(\boldsymbol{x}, \boldsymbol{y})],
\end{equation}
where $\boldsymbol{x}$ and $\boldsymbol{y}$ are the dimensionless image and source positions in Eq.~\eqref{position}, $M^{z}_{L} \equiv M_L\left(1+z_L\right)$ is the redshifted lens mass at redshift $z_{L}$, and $t(\boldsymbol{x},\boldsymbol{y})$ is the time delay at $\boldsymbol{x}$ due to the lensing potential and different geometrical path length.

In the strong lensing regime, where the lens size is galaxy or galaxy cluster scale,
the time delays between lensed images are typically longer than the durations of GW signals from compact binary coalescences.
Consequently, geometrical optics is valid for GWs, and the amplification factor in Eq.~\eqref{af} can be approximated as
\begin{equation}
\label{Fgeo}
F_{\rm geo}(f,\boldsymbol{\theta_{l}}) = \sum_{j}|\mu_{j}|^{0.5}  \exp{i f t_{j} -i \pi \phi_{n,j}},
\end{equation}
where $\boldsymbol{\theta_{l}} = \{\mu_j, t_j, \phi_{n,j} \}$ are lens parameters and $\mu_{j}$, $t_{j}$, and $\phi_{n,j}$ are respectively the magnification factor, time delay, and Morse factor for the $j^\mathrm{th}$ image~\citep{wang1996gravitational,takahashi2003wave,dai2017waveforms}.
At a given image position, $\mu_j$ and $t_j$ are continuous quantities, whereas $\phi_n$ takes on a discrete value among 0, 0.5, and 1. 
These correspond to the image type (Type I, II, or III), which is a discrete topological classification determined by the nature of the stationary point on the time delay surface.

In the microlensing regime, where the lensing objects are comparable in size to the GW wavelength, wave optics becomes important.
Microlensed GWs typically appear as single distorted signals due to short time delays between multiple microimages, with interference patterns causing frequency-dependent modulation in the signals~\citep{deguchi1986diffraction,nakamura1998gravitational,takahashi2003wave}.
To generate microlensed waveforms, the amplification factors are computed by solving the diffraction integral in Eq.~\eqref{af}.
Often, an isolated point mass is employed for the microlens model, characterized by the lensing potential $\psi(\boldsymbol{x}) = \ln{|\boldsymbol{x}|}$, as its integral solution is analytic. 

\begin{figure}[t]
    \centering
\includegraphics[width=1.0\linewidth]{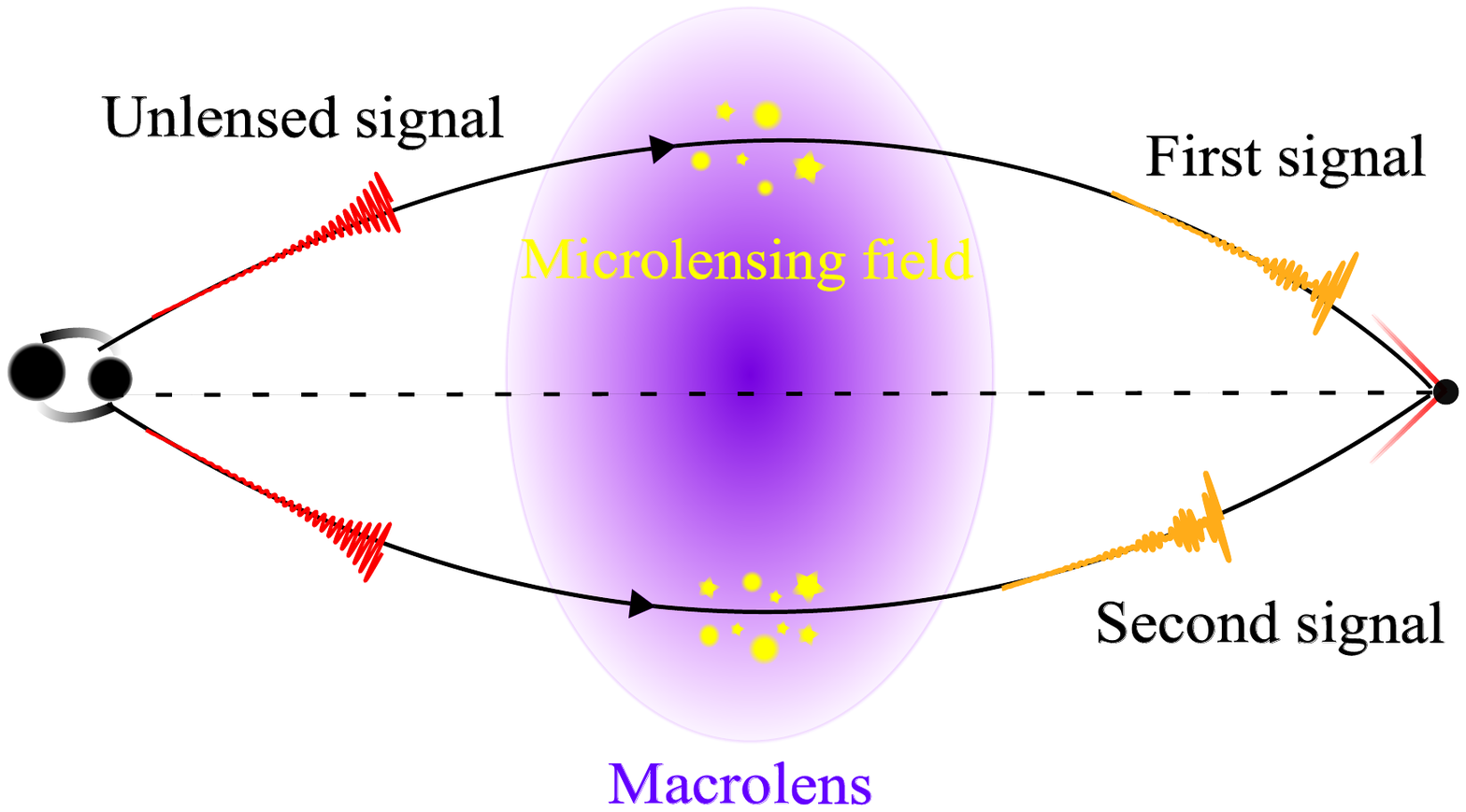}
    \caption{A lensing system creates two lensed GW signals from an unlensed signal of a BBH, where the lens consists of a macrolens (galaxy) and  microlensing fields (stellar-mass objects).
    The presence of microlensing fields can produce extra interference patterns on the macroimages.
    Note that microlensing fields at each macroimage are distinct, as a stellar distribution in a galaxy is not uniform meaning each image undergoes a different frequency modulation.}
    \label{lensing_schematic}
\end{figure}

In a more realistic lensing scenario, microlens candidates are predominantly located within substructures of galaxies or in the intercluster medium embedded within galaxy clusters.
In addition, the density of stellar fields in these substructures can be sufficiently high for GWs to undergo multiple-plane lensing~\citep{diego2019observational,PhysRevD.101.123512,meena2020gravitational,mishra2021gravitational}.
Figure~\ref{lensing_schematic} shows the schematics of the lensing system with a macrolens containing fields of microlenses.
A GW propagating through the object is first strongly lensed by the macrolens, splitting into multiple macroimages.
Each macroimage is then further microlensed by a stellar field located nearby.

In this case, the total lensing effect on a GW signal from microlenses embedded within a macrolens can be estimated using Eq.~\eqref{af} with an appropriate time delay function $t(\boldsymbol{x},\boldsymbol{y})$.
The time delay function must be established considering the macrolens contributions~\footnote{We assume a singular isothermal ellipsoid (SIE)~\citep{kormann1994isothermal} as our macrolens model.} to the microlensing effects, which can be parametrized by convergence ($\kappa$) and shear ($\gamma$)~\citep{schneider2006gravitational}.

Given this setup, the time delay function for $N$ point masses near a macroimage located at $\boldsymbol{x}= (x_{1},x_{2})$ is given by~\citep{zheng2022improved,shan2023microlensing,Shan:2024min}
\begin{equation}\label{microfermat}
\begin{split}
t\left(\boldsymbol{x}, \boldsymbol{x}^m, \boldsymbol{y}_{\rm macro}=0\right)=&\frac{k}{2}\left((1-\kappa+\gamma) x_1^2+(1-\kappa-\gamma) x_2^2\right)\\
&-\left[\frac{k}{2} \sum_m^N \ln \left(\boldsymbol{x}^m-\boldsymbol{x}\right)^2+k \phi_{-}(\boldsymbol{x})\right],
\end{split}
\end{equation}
where $\boldsymbol{x}^m$ is the position of the $m^{\mathrm{th}}$ point mass in the stellar field, and $k=4Gm^{z}_{\rm L}/c^{3}$, where $m_{\rm L}$ indicates average microlens mass. 
$\phi_{-}$ is the negative mass sheet term to offset the effects from the microlens masses, which does not change the total convergence.
In Eq.~\eqref{microfermat}, the macrolensing contribution is obtained through an approximation expansion, and the center of the expansion can be chosen manually.
For convenience, we choose the macroimage point as the expansion center, such that $\boldsymbol{y}_{\rm macro} = 0$.
By inserting Eq.~\eqref{microfermat} into Eq.~\eqref{af}, one can calculate the amplification factors $F(f)$ of a stellar field embedded in a macrolens~\citep{Shan:2022xfx}.
\subsection{Data generation and joint parameter estimation} \label{sec:jpe}
We use \textsc{Pycbc}~\citep{alex_nitz_2024_10473621} to simulate 100 mock O4-like BBHs using the \textsc{IMRPhenomPv2} waveform model~\citep{hannam2014simple}. 
Next, we consider two lensing hypotheses to create two lensed signal pairs for each BBH.
1) $\mathcal{H}_{\rm SL}$: a BBH is strongly lensed by a single macrolens producing two images, and 2) $\mathcal{H}_{\rm MLSL}$: a BBH is strongly lensed by the same macrolens in $\mathcal{H}_{\rm SL}$ and the two macroimages are further microlensed by stellar fields embedded in the macrolens.
Thus, we generate a total of 100 pairs of strongly lensed GW signals ($h^{\rm inj}_{\rm SL, 1}, h^{\rm inj}_{\rm SL, 2}$) and 100 pairs of strongly lensed and microlensed GW signals ($h^{\rm inj}_{\rm MLSL, 1}, h^{\rm inj}_{\rm MLSL, 2}$) as injections. 
Note that the lensing effect in $\mathcal{H}_{\rm SL}$ is  well described by $F_{\rm geo}(f)$ in Eq.~\eqref{Fgeo}, while, for $\mathcal{H}_{\rm MLSL}$ the amplification factor corresponds to $F(f)$ in Eq.~\eqref{af} with Eq.~\eqref{microfermat}.
Table~\ref{tab:params} summarizes the main parameter distributions and ranges used to generate strongly lensed and microlensed waveforms.
Further details on the data generation process can be found in Appendix~\ref{app:mock_data}.

\begin{table}[th]
\begin{tabular*}{\linewidth}
{@{\extracolsep{\fill}} llll}
\toprule
\textbf{Parameter} &
\textbf{Distribution or Model} & \textbf{Range} \\
\hline
$m_{1}$ & \texttt{Power Law + peak} & [5, 85]$M_{\odot}$  \\
$m_{2}$ & \texttt{Uniform} & [5, $m_{1}$]$M_{\odot}$\\
$m_{\rm L}$ (stellar) & \texttt{Salpeter IMF} & [0.1, 1.5]$M_{\odot}$\\
$m_{\rm L}$ (remnant)& \cite{spera2015mass} & [0.1, 28]$M_{\odot}$ \\
$\kappa_{*}$ & \texttt{Sérsic profile} & - \\
$N(\kappa_{*},m_{L})$ & - & [$10^3$, $10^6$]  \\ 
$\kappa$, $\gamma$ & \texttt{SIE} ($q,\sigma_{v},y$) & -  \\
        \hline
        \hline
    \end{tabular*}
    \caption{The assumptions for component masses ($m_1$, $m_2$),   macrolens, and microlens parameters employed to simulate strongly lensed and microlensed GW signals.
    For each lensing system, the number of microlenses ($N$) is determined by the microlensing convergence ($\kappa_{*}$) and the microlens mass ($m_{\rm L}$), representing contributions from stellar and remnant components.
    The convergence ($\kappa)$ and shear ($\gamma$) of macrolens are derived from the SIE model, characterized by three parameters: the axis ratio ($q$), the velocity dispersion ($\sigma_{v}$), and the dimensionless source position ($y$). Details on the sampling of each parameter are provided in Appendix~\ref{app:mock_data}.
    \label{tab:params}}
\end{table}
We inject all the lensed signals into the aLIGO-Hanford, aLIGO-Livingston, and Advanced Virgo with stationary Gaussian noises $n$ simulated using the O4 design sensitivity PSD~\citep{abbott2020prospects}.~\footnote{The PSD data were obtained from the LIGO Document Control Center (DCC) under document LIGO-T2000012, available at https://dcc.ligo.org/LIGO-T2000012/public.}
Also, when assessing future prospects, we use the simulated PSD of LIGO O5~\footnote{The expected O5 PSD data for each detector used for joint parameter estimations can be found from $\textsc{Pycbc.psd}$ package~\citep{alex_nitz_2022_6912865}.}~\citep{abbott2020prospects}.
We only consider signals with a network signal-to-noise ratio (SNRs) higher than the threshold value $\rho_{\rm net} =12$, following \cite{shan2023microlensing}.

To analyze the 200 pairs of lensed GW signals, we use the \textsc{Golum} pipeline~\citep{janquart2021fast,janquart2023return}.
The joint likelihood function in the pipeline assumes that the two input GW signals are from the same source and solely strongly lensed, implying that microlensing effects are unaccounted for.

We obtain maximum likelihood waveforms of the first lensed signals ($h^{\rm maxL}_{\rm SL,1}$) from the joint parameter estimations (JPEs) carried out for 100 pairs of strongly lensed signals.
In addition, we extract the maximum likelihood values of the lensing parameters, including relative magnification factors ($\mu_{\rm rel}$), differences between arrival times ($\Delta t$), and differences between Morse factors ($\Delta \phi_{n}$).
One can calculate the maximum likelihood waveforms of the second lensed GW signals ($h^{\rm{maxL}}_{\rm SL,2}$) by converting luminosity distances, coalescence time, and Morse factor as follows,
\begin{eqnarray}
\label{onetotwo}
D^{\rm{maxL}}_{L}|_{\rm{SL,2}}&= D^{\rm{maxL}}_{L}|_{\rm{SL,1}}& \times \sqrt{\mu_{\rm{rel}}^{\rm{maxL}}}, \nonumber \\
 t^{\rm{maxL}}_{c}|_{\rm{SL,2}} &= t^{\rm{maxL}}_{c}|_{\rm{SL,1}} &+ \Delta t^{\rm{maxL}},  \nonumber \\ 
 \phi^{\rm{maxL}}_{n}|_{\rm{SL,2}}&= \phi^{\rm{maxL}}_{n}|_{\rm{SL,1}}& + \Delta \phi_{n}^{\rm{maxL}}.
\end{eqnarray}
It is important to note that even in cases where lensed waveforms exhibit additional phase modulations due to significant eccentricity, precessing spins, or higher-order modes (HMs)~\citep{janquart2021identification,PhysRevD.103.064047,vijaykumar2023detection}, Eq.~\ref{onetotwo} is valid provided that JPE employs waveform models that incorporate these features. 
We underline that a JPE analyzes two lensed signals simultaneously, enabling accurate inference of lens parameters, such as the Morse factors and their difference~\citep{janquart2021identification}.
Then, $h^{\rm{maxL}}_{\rm SL,2}$ can be obtained from $h^{\rm{maxL}}_{\rm SL,1}$ in the frequency domain.

Similarly, the maximum likelihood waveforms of pairs of microlensed signals ($h^{\rm maxL}_{\rm MLSL,1}$, $h^{\rm maxL}_{\rm MLSL,2}$) can be obtained.
Note that $h^{\rm maxL}_{\rm MLSL}$ has a different waveform morphology from $h^{\rm inj}_{\rm MLSL}$ as the GW templates in the pipeline used are solely strongly lensed ones, which should lead to larger residuals. 
\section{Residual test} \label{sec:residualtest}
Since the adopted pipeline does not account for microlensing effects, the discrepancies between $h^{\rm maxL}_{\rm MLSL}$ and $h^{\rm inj}_{\rm MLSL}$ should typically be more significant than those between $h^{\rm maxL}_{\rm SL}$ and $h^{\rm inj}_{\rm SL}$.
To statistically quantify these differences between the maximum likelihood waveforms retrieved under the two hypotheses, we investigate the residuals ($r \equiv d-h$), which correspond to the strain that remains after subtracting the signal model ($h$) from the detected data ($d$).
If the true GW signal embedded in the detected data perfectly matches the signal model, the residual will be consistent with pure detector noise~\footnote{There are instances when non-Gaussian features, such as glitches, occur in the signal at the detected time. In such cases, the residuals do not align with the normal detector noise observed at different times.}.

We assume that the data, containing a lensed GW signal, in the $i^{\rm th}$ detector can be written as
\begin{equation}
d^{i} =h^{i}_{\rm SL}+ \mathrm{d}h^{i} + n^{i},
\end{equation}
where $h_{\rm SL}$ is the pure strongly lensed signal that \emph{does not include the effects of microlensing},
$\mathrm{d}h$ represents the frequency-dependent residual induced by microlensing effects (i.e. $dh=0$ in the $\mathcal{H}_{\mathrm{SL}}$ case), and $n$ is the detector noise.

Under hypothesis $\mathcal{H_{\rm SL}}$, the lensing effects are governed by $F_{\rm geo}$ and the waveform morphology does not change, which implies that the $dh$ terms vanish.
Thus, the residual in the $i^{\rm th}$ detector is given by
\begin{align}
\label{resSL}
    r^{i}|\mathcal{H}_{\rm SL} &\equiv \lim_{\rm{d}\it{h^i}\rightarrow \mathbf{0}} \left (d^{i}-h^{i,\rm{maxL}}\right) \nonumber  \\ &=  h^{i}_{\rm SL} + n^{i} - h^{i,\rm{maxL}} \nonumber \\ &=\delta h^{i}_{\rm SL} + n^{i},
\end{align}
where $\delta h_{\rm SL}$ is the difference between the true lensed signal and the maximum likelihood waveform recovered under $\mathcal{H}_{\rm SL}$. If JPE recovers the waveform parameters accurately, this difference will be small (i.e. $\delta h_{\rm SL}\rightarrow 0$).
In the optimal case, the remaining residual strain ($r_{\rm SL}$) in each detector should be consistent with instrumental noise ($n_i$).

\begin{figure}[t]
    \centering
\includegraphics[width=1.0\linewidth]{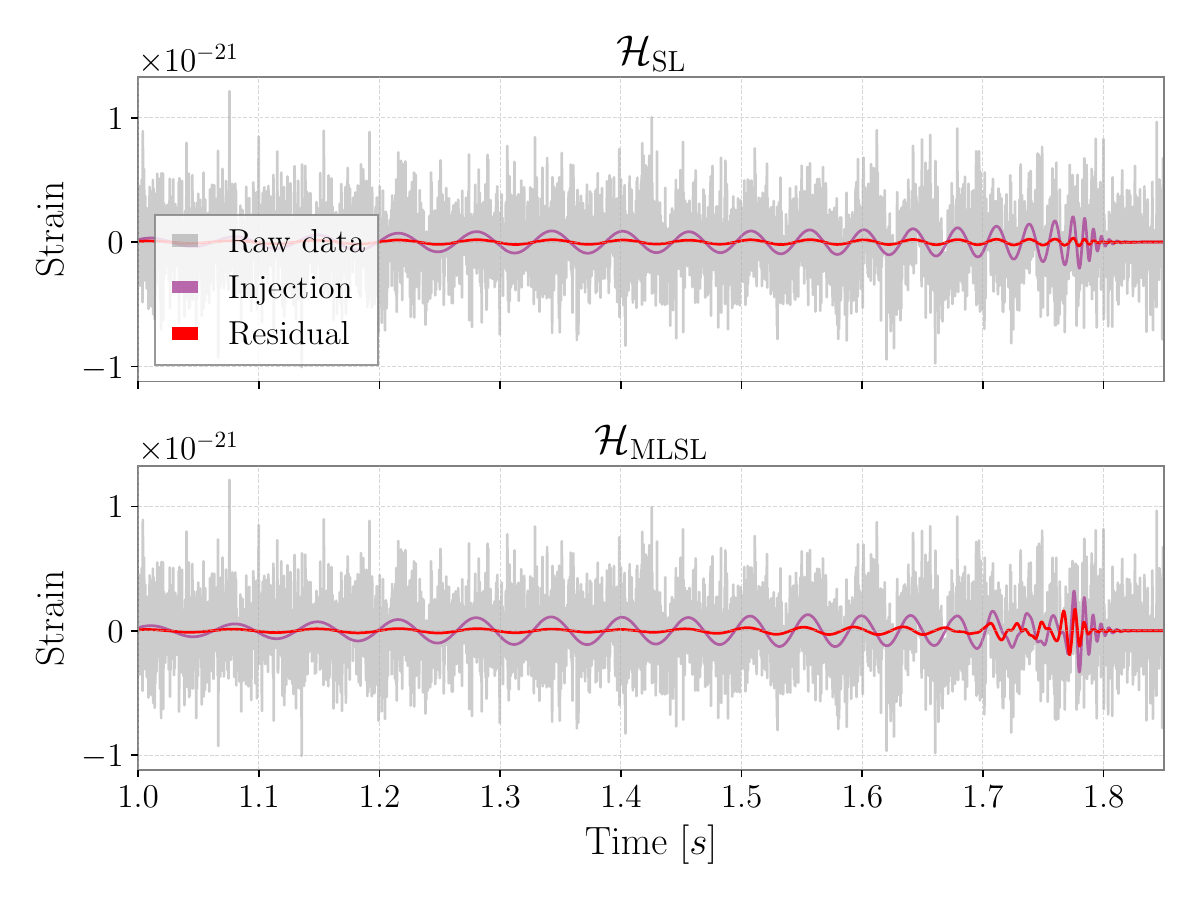}
    \caption{Representation of the time domain strains of raw data (grey solid lines), injected waveforms (purple solid lines), and residuals (red solid lines) for a representative BBH event in the LIGO-Handford detector, assuming $\mathcal{H}_{\rm SL}$ and $\mathcal{H}_{\rm MLSL}$ in the top and bottom panel, respectively.
    We assume the same BBH and macrolens parameters for the two injected waveforms.
    The residual from $\mathcal{H}_{\rm MLSL}$ is more significant than that from $\mathcal{H}_{\rm SL}$.}
    \label{residual_strain}
\end{figure}

On the other hand,  $\rm{d}\it{h}$ has a non-zero value under hypothesis $\mathcal{H_{\rm MLSL}}$.
In this case the residual can be written as
\begin{align}
\label{resMLSL}
    r^{i}|\mathcal{H}_{\rm MLSL} &\equiv d^{i}-h^{i,\rm{maxL}} \nonumber  \\ &=  h^{i}_{\rm SL} + \mathrm{d}h^{i} + n^{i} - h^{i,\rm{maxL}} \nonumber \\ &=\delta h^{i}_{\rm MLSL} +  n^{i},
\end{align}
where $\delta h_{\rm MLSL}$ contains both the difference between the strongly lensed signal part and the maximum likelihood waveform model {\em and} the extra microlensing contribution, $dh$.
Unlike $\delta h_{\rm SL}$, $\delta h_{\rm MLSL}$ should be non-zero since the JPE pipeline analyzes signals with waveform templates assuming $\mathcal{H}_{\rm SL}$.
As a result, the residual strain obtained from microlensed signals ($r_{\rm MLSL}$) should be different from $n_{i}$.
A representation of the residuals expected under each hypothesis is given in Figure~\ref{residual_strain}.

To evaluate how much the residuals deviate from pure detector noise (denoted by $\mathcal{H}_{\rm N}$) under a given signal hypothesis $\mathcal{H}_{\rm X}$ (where $\rm X$ is either SL or MLSL), we analyze the correlation between residuals in different detectors.
We introduce a factor $\Lambda$ representing this quantity.\footnote{In \cite{chakraborty2024mu}, the cross-correlations of residuals—defined as the deviations between observed microlensed strain data and the best-fit unlensed templates—are utilized to search for microlensing signatures in unlensed GW observations. Based on \cite{chakraborty2024glance}, the authors compute these cross-correlations by introducing a time delay $\tau$ to slide the residual in one detector relative to the residual in another detector. This approach assesses how well the two signals align at various time delays. We adapt the cross-correlation technique in our analysis by calculating the inner product of the residuals from each detector to obtain the $\Lambda$ value, which quantifies the alignment between the two signals.}
For strain data from the $i^{\rm th}$ and $j^{\rm th}$ detectors, containing residuals $r^{i}$ and $r^{j}$ respectively, $\Lambda$ is defined as
\begin{equation}
\label{lambda}
\Lambda_{\mathcal{H}_{\rm {X}}} = 
\exp\bigg[\frac{1}{2} 
\big< r^{i}(t)_{\mathcal{H}_{\rm X}},
r^{j}(t+\tau)_{\mathcal{H}_{\rm X}} \big> \bigg],
\end{equation}
where $\tau$ is the time delay between the arrival times of the GW at the $i^{\rm th}$ and $j^{\rm th}$ detectors, and the inner product $<.,.>$ is given by
\begin{equation}
\label{inner_product}
\langle\mathbf{a}, \mathbf{b}\rangle=2 \int_0^{\infty} \frac{\tilde{a}(f) \tilde{b}^*(f)+\tilde{a}^*(f) \tilde{b}(f)}{S_n(f)} \mathrm{d} f~,
\end{equation}
in the frequency domain, where $^*$ denotes the complex conjugate and $S_{n}$ is the one-sided power spectral density (PSD).

When we calculate $\Lambda$ using the residuals, we do not use a time window that covers the entire signal duration.
Instead, we adopt time windows of the form
\begin{equation}
t \in [{t_c}- \alpha t_s+\Delta t, t_c +\Delta t] ,
\label{time_window}
\end{equation}
where $t_{c}$ is the merger time, $\alpha$ is a constant to adjust the time window size, $\Delta t$ denotes the time interval that sufficiently covers the ringdown phase (we use $\Delta t = 0.2s$), and $t_{s}$ is the signal duration based on the JPE results.
In what follows a window with $\alpha = 0.25$ was chosen since it gives the optimal results for our residual analysis.
Further discussion on the time window efficiency will be presented in Sec.~\ref{sec:4.3}.

Recall that the residual $r$ is decomposed into the model mismatch $\delta h$ and detector noise $n$ components.
When evaluating Eq.~\eqref{lambda}, the $\delta h$ component contributes directly to the inner product, as it is coherent across the segments with a correct $\tau$.
However, the $n$ components are uncorrelated between detectors due to their stochastic nature, so their contributions tend to average out to zero.
When the value of $\delta h$ is significant, $\Lambda$ tends to have a larger value.

The $\Lambda$ distributions obtained from each hypothesis can be differentiated if their corresponding residuals are sufficiently distinctive.
In this work, the $\Lambda_{\mathcal{H}_{\rm SL}}$ distribution effectively serves as a background, allowing us to determine which hypothesis is favored for a given microlensed GW event based on where the computed value of $\Lambda$ lies within this background distribution.
We assume that a microlensed event favors $\mathcal{H}_{\rm MLSL}$ if the event has a $\Lambda_{\mathcal{H}_{\rm MLSL}}$ value larger than the upper bound of the 2$\sigma$ 
credible interval \footnote{Note that all one-dimensional credible intervals computed in this paper are central, i.e. with equal probability below and above the lower and upper bounds.} (C.I.) of the $\Lambda_{\mathcal{H}_{\rm SL}}$ distribution.
It is important to note that while the Virgo detector is included in the PE process, it is not used for calculating $\Lambda$ in Eq.~\ref{lambda}, as the noise level of the Virgo detector is generally greater than that of the LIGO detectors. 
Of course, a detector network with more detectors is beneficial due to the increase in SNR. 
However, in the context of the residual test, if a detector has a high noise level, the $\Lambda$ values, including the cross-correlation terms from that detector as shown in Eq.~\eqref{lambda}, have a broader distribution.
This can lead to some microlensed events being unidentified based on our criterion.

\begin{figure*}[t]
    \centering
    \includegraphics[width=1.0\linewidth]{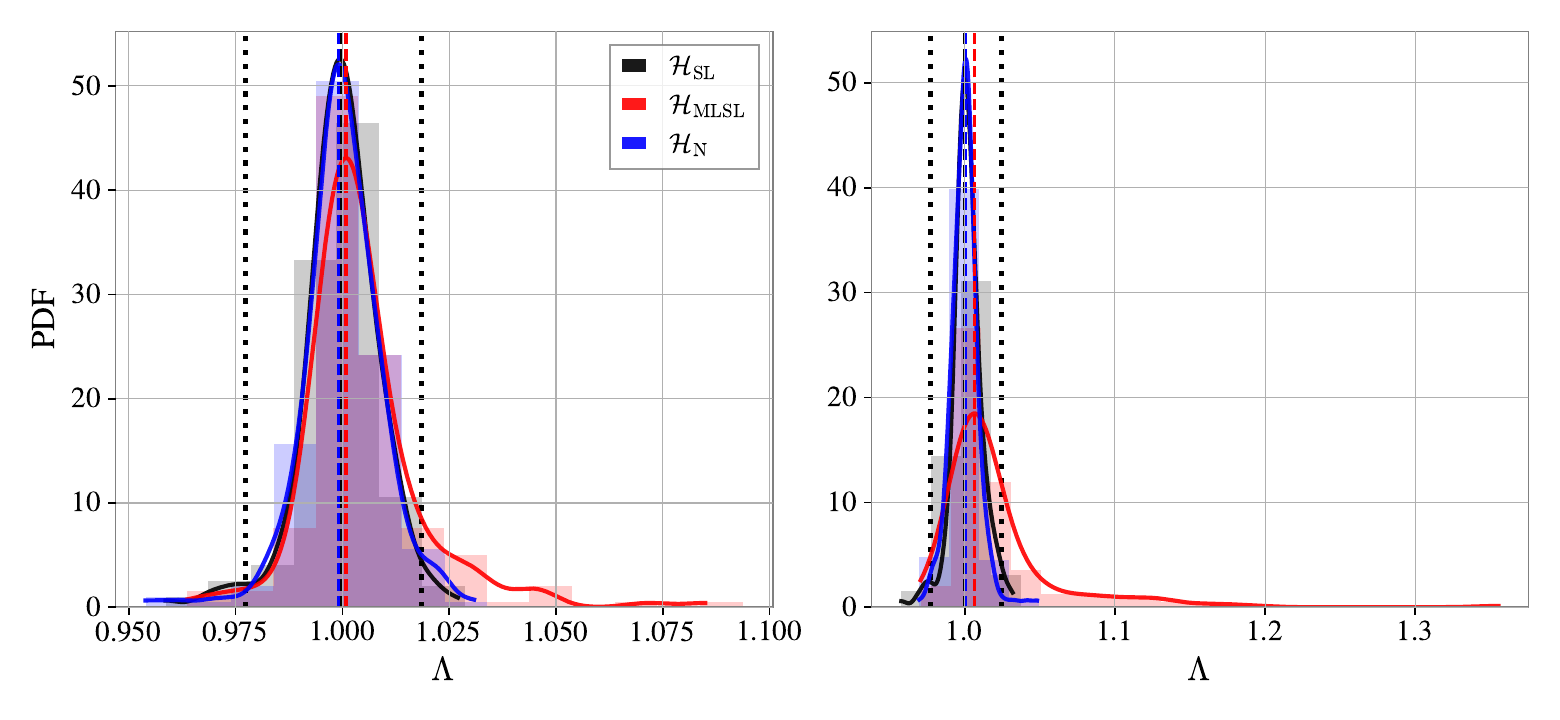}

    \caption{\textit{Left panel}: The $\Lambda$ distributions and their corresponding Gaussian KDEs for each of 200 $\boldsymbol{\rm N}$ (blue), $\boldsymbol{r}_{\rm SL}$(black), and $\boldsymbol{r}_{\rm MLSL}$ (red) samples obtained using the O4 sensitivity.
    $\boldsymbol{r}_{\rm SL}$ and $\boldsymbol{r}_{\rm MLSL}$ have the same underlying noise strain.
    Vertical dashed lines denote the peak values of the Gaussian KDEs, and dotted black lines are 2$\sigma$ C.I. for $\Lambda_{\mathcal{H}_{\rm SL}}$, representing the threshold for identifying microlensed events in our residual test.
    \textit{Right panel}: The results obtained by employing the O5 sensitivity.
    The number of events with $\Lambda$ values located above the right dotted black line increases compared to the O4 results, indicating that the microlensing detectability is improved.
   }\label{lambda_distribution}
\end{figure*}
\section{Results} \label{sec:result}
\subsection{Analysis using realistic population-based events}
From the JPE conducted for the two independent sets of 100 pairs, created under the scenarios of strong lensing and superimposed microlensing on strong lensing (see Sec.~\ref{sec:jpe}), we obtain maximum likelihood waveforms (i.e. 200 $h^{\rm maxL}_{\rm{SL}}$ and 200 $h^{\rm maxL}_{\rm{ MLSL}}$)
By using Eqs.~\eqref{resSL}-\eqref{lambda}, we compute the values of $\Lambda_\mathcal{H_{\rm SL}}$ from 200 $r_{\rm SL}$ and $\Lambda_\mathcal{H_{\rm MLSL}}$ from 200 $r_{\rm MLSL}$, where the instrumental noise $n^{i}$ in both $r^{i}_{\rm SL}$ and $r^{i}_{\rm MLSL}$ are the same.
Additionally, we generate another set of 200 pure Gaussian noise segments ($\rm N$) for each of the two detectors, based on the PSDs employed to generate the data, resulting in corresponding $\Lambda_{\mathcal{H}_{\boldsymbol{\rm N}}}$ values.

Since we assume that the detector noise is stationary and Gaussian, the $\Lambda_{\mathcal{H}_{\boldsymbol{\rm SL}}}$ distribution is expected to be close to the $\Lambda_{\mathcal{H}_{\boldsymbol{\rm N}}}$ distribution, which is a normal distribution with a mean of 1 and unit variance.
In contrast, the distribution for $\Lambda_{\mathcal{H}_{\boldsymbol{\rm MLSL}}}$ deviates from the distribution for $\Lambda_{\mathcal{H}_{\boldsymbol{\rm N}}}$, indicating that the additional microlensing-induced terms in Eq.~\eqref{resMLSL} are more significant than $\delta h_{\rm SL}$. 

Figure~\ref{lambda_distribution} shows the $\Lambda$ distributions for the residuals and Gaussian noises and their corresponding Gaussian kernel density estimates (KDEs).
To determine which event supports $\mathcal{H}_{\rm MLSL}$ based on the $\Lambda$ distributions, we consider 2$\sigma$ credible interval (C.I.) of the $\Lambda_{\mathcal{H}_{\boldsymbol{\rm SL}}}$ distribution. 
If an event has a $\Lambda$ value larger than the upper bound of the 2$\sigma$ C.I, the event favors $\mathcal{H}_{\rm MLSL}$ and it is considered a candidate for being both strongly lensed and microlensed.

For O4 sensitivity, 11\% of 200 microlensed events favor $\mathcal{H}_{\rm MLSL}$.
One of the reasons for this somewhat low percentage is that most microlensed events created for our realistic microlensed population have small mismatch (MM) values (MM $\leq 0.03$ ) compared to $h_{\rm SL}$ (see App.~\ref{app:mock_data} for details), where MM is defined as
\begin{align}
\textrm{MM} &\equiv 1 - \mathcal{M}\left(h_{\rm MLSL}, h_{\rm SL}\right) \nonumber \\
&= 1- \max_{\phi_0, t_0} 
\frac{\left\langle h_{\rm MLSL}, h_{\rm SL} \right\rangle}
{\sqrt{\left\langle h_{\rm MLSL}, h_{\rm MLSL} \right\rangle 
\left\langle h_{\rm SL}, h_{\rm SL} \right\rangle}}.
\end{align}
Such events are also expected to be hard to detect using more complete microlensing analysis pipelines, such as \textsc{Gravelamps}~\citep{wright2022gravelamps}.
We will discuss in Sec.~\ref{sec:4.2} how the results change for the large mismatch events that are expected to be detectable.

In addition to considering the O4 sensitivity, we also conduct the same analyses again for LIGO at the O5 sensitivity to determine how much our residual test to detect microlensing signatures is improved. This will be the case if (1) the $\delta h_{\rm SL}$ term in Eq.~\ref{resSL} is minimized and (2) the noise level is reduced to maximize the contribution of $dh$ in Eq.~\ref{resMLSL} to $\Lambda$.
As expected, the percentage of microlensed events favoring $\mathcal{H}_{\rm MLSL}$ increases to 21.5\% at O5 sensitivity, for the same population of strongly lensed and microlensed events.

\begin{figure*}[t]
    \centering
    \subfigure[O4 sensitivity]{\includegraphics[width=0.75\linewidth]{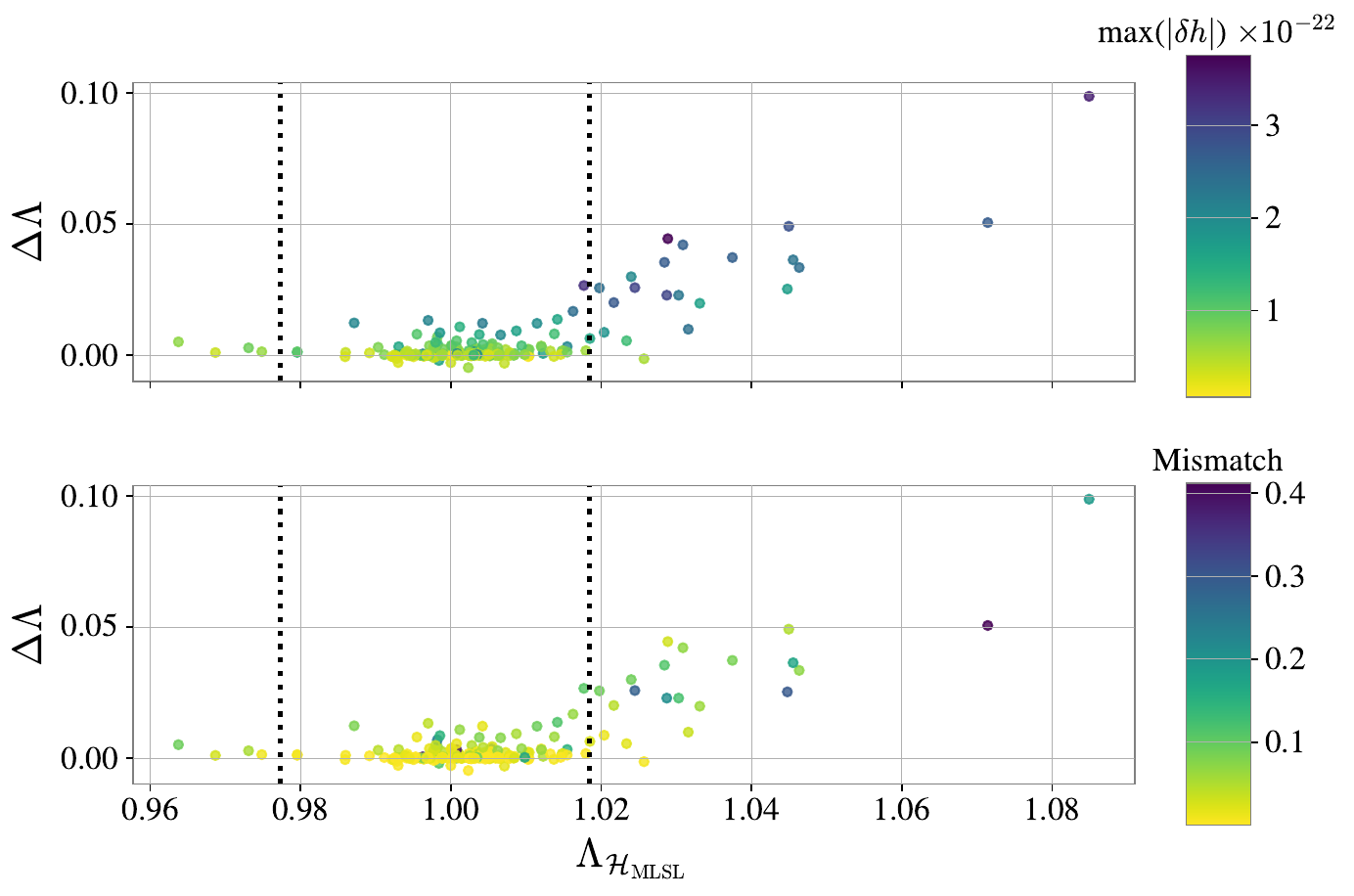}}
    \subfigure[O5 sensitivity]{\includegraphics[width=0.75\linewidth]{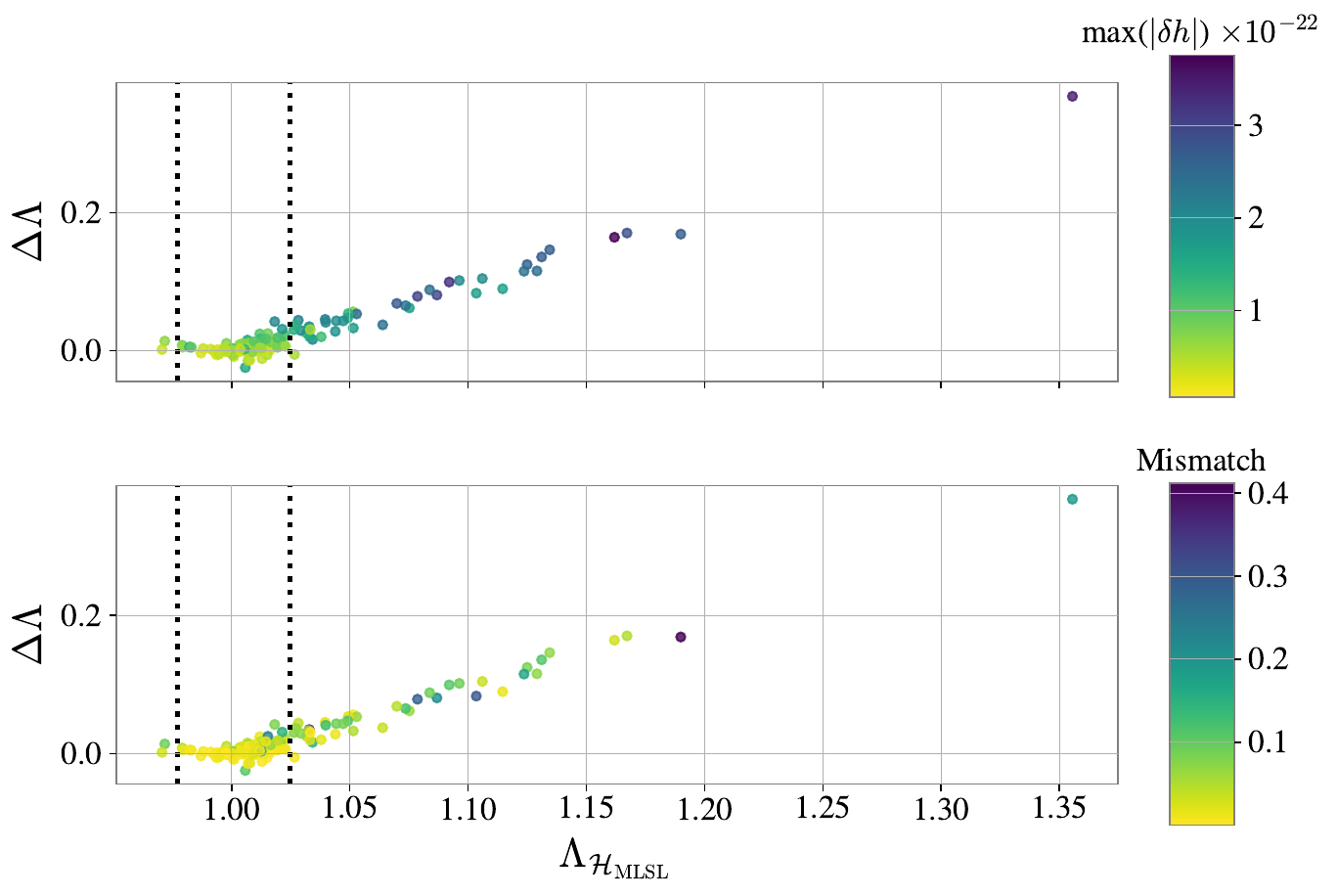}}
    \caption{Scatter plots showing the correlation between $\Lambda_{\mathcal{H}_{\rm MLSL}}$ and $\Delta \Lambda$ for (a) O4 and (b) O5 sensitivities, color-coded based on max$(|\delta h|)$ and mismatch, denoting the maximum strain of the pure residual and mismatch between $h_{\rm SL}$ and $h_{\rm MLSL}$ computed by averaging the values obtained from aLIGO-Hanford and aLIGO-Livingston, respectively.
    The $2\sigma$ C.I. for $\Lambda_{\mathcal{H}_{\rm SL}}$ in Fig.~\ref{lambda_distribution} are plotted in dotted black lines.
    Remind that only events located above the right dotted line can be identified as microlensed.
    A larger difference between $h_{\rm SL}$ and $h_{\rm MLSL}$ corresponds to a larger $\Lambda_{\mathcal{H}_{\rm MLSL}}$ and a larger $\Delta \Lambda$.
    High mismatch and high max$(|\delta h|)$ values are consistent with high $\Delta \Lambda$, but the statistical fluctuation of noise can prevent a microlensed event with positive $\Delta \Lambda$ from being identified as microlensed.}
    \label{mismatch}
\end{figure*}

To verify that more significantly microlensed events have larger $\Lambda$ values, we calculate the differences between the $\Lambda$ values (i.e. $\Delta \Lambda \equiv \Lambda_{\mathcal{H}_{\rm MLSL}} - \Lambda_{\mathcal{H}_{\rm SL}}$) obtained from the residuals $r_{\rm SL}$ and $r_{\rm MLSL}$ for individual BBH events; this allows us to investigate their correlations with MM values and with the maximum strain of $\delta h_{\rm MLSL}$, representing the amplitude of the waveform left in the data after subtracting the strongly lensed model.
 $\Delta \Lambda > 0$ indicates that $r_{\rm MLSL}$ is larger than $r_{\rm SL}$.

As shown in Figure~\ref{mismatch},
microlensed events with a higher value of $\delta h$ are located in the high $\Delta \Lambda$ region, which is expected given the definition of $\Lambda$ in Eq.~\eqref{lambda}.
Similarly, microlensed events with relatively high MM values tend to have high $\Delta \Lambda$ values, although the correlation between $\Delta \Lambda$ and $\delta h$ appears stronger than that between $\Delta \Lambda$ and the MM.
However, in the O4 case most of the high-MM events are located within the $2\sigma$ C.I. of the $\Lambda_{\mathcal{H}_{\rm SL}}$ (black dotted lines), as the noise power is relatively high compared to the residual power; consequently the statistical fluctuations of $\Lambda_{\mathcal{H}_{\rm SL}}$ tend to be broader than those of $\Delta \Lambda$.
In contrast, in the O5 case,  the reduced noise level results in most events with high MM values being located above the upper bound of the $2\sigma$ C.I. for $\Lambda_{\mathcal{H}_{\rm SL}}$.

It is important to note that a high MM value does not necessarily indicate a large $\delta h$. 
If a significantly distorted signal is weak (i.e. of low SNR), the absolute value of $\delta h$ remains low despite the high MM. On the other hand, even if a weakly distorted signal is detected, the time delay between the two microlensed images can be very short, resulting in a repeated merger part of the waveform.
In such cases, the recovered best-fit waveform may not fully subtract the repeated part, leading to a high $\delta h$ value, which corresponds to a high value of $\Delta \Lambda$.

\subsection{Analysis using high-mismatch events}\label{sec:4.2}
\begin{figure*}[t]
    \centering
\includegraphics[width=1.0\linewidth]{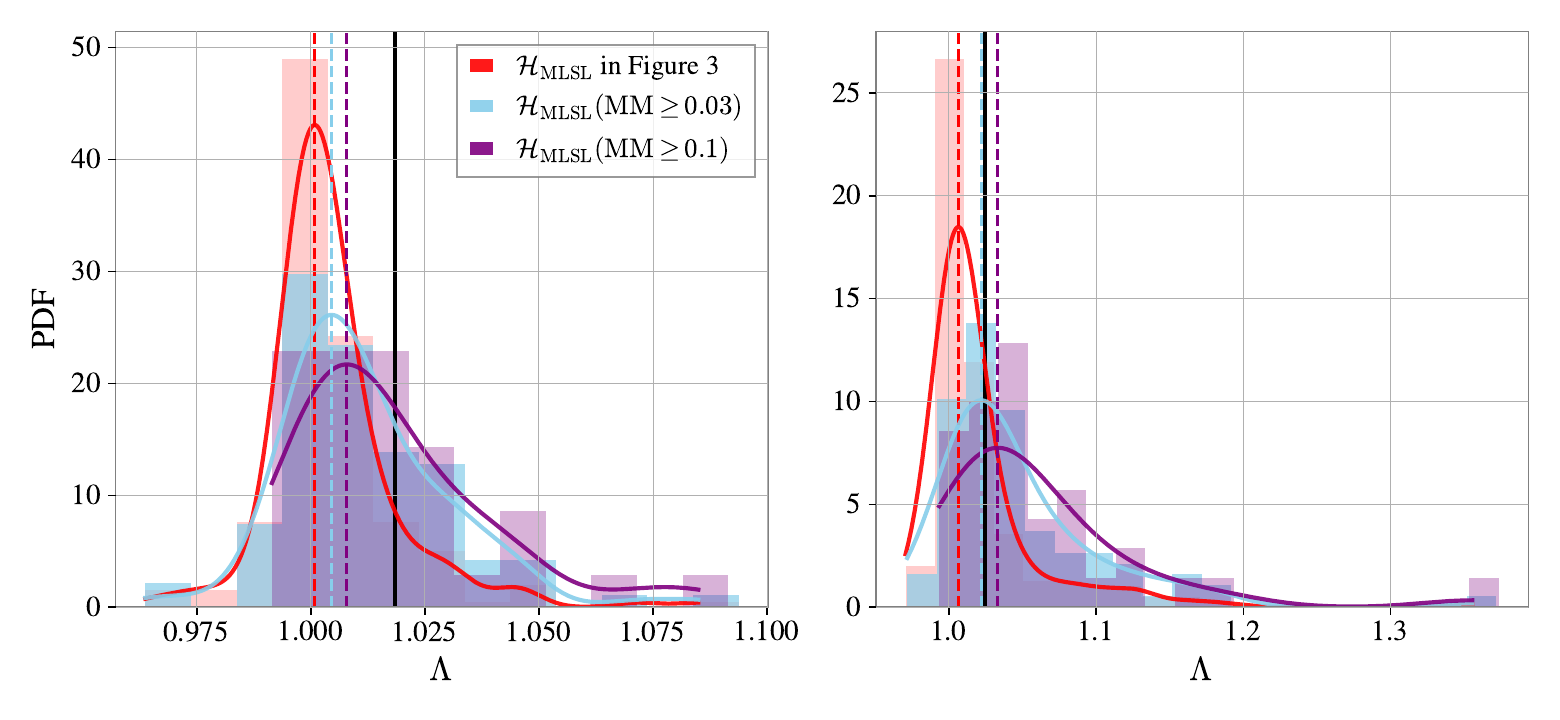}
    \caption{The same configuration as Fig.~\ref{lambda_distribution}, but now showing $\Lambda$ distributions for realistic population-based microlensed events (red), events with MM $\geq 0.03$ (cyan) and with MM $\geq 0.1$ (purple).
    The solid black line is the upper bound of the 2$\sigma$ C.I. for $\Lambda_{\mathcal{H_{\rm SL}}}$ that was shown in Fig.~\ref{lambda_distribution}.
    The $\Lambda$ distributions for the microlensed events with higher mismatch are further shifted and posititively skewed compared to those for realistic population-based microlensed events, indicating that the residual test more easily identifies more significantly distorted signals.
    Correspondingly, the ratio of events identified as microlensed increases with higher mismatch values.}
    \label{high_mismatch}
\end{figure*}
As aforementioned, most of the simulated $h_{\rm MLSL}$ have small mismatch values, which aligns with expectations from a realistic microlensing population~\citep{mishra2021gravitational}.
Therefore, we also focus on microlensed events with MM $\geq 0.03$ and MM $\geq 0.1$~\footnote{Mismatch values are computed by averaging the two mismatch values obtained using $h_{\rm SL}$ and $h_{\rm MLSL}$ detected in the aLIGO-Hanford and aLIGO-Livingston detectors.}, which are expected to be detectable with the current pipeline, to better assess the effectiveness of the residual test.
To obtain more robust results, we also create additional events with high mismatch values ($\rm{MM} \geq 0.3$) by artificially enhancing the microlensing effects on strongly lensed GW signals (see Appendix~\ref{app:mock_data}).

Figure~\ref{high_mismatch} compares the $\Lambda$ distributions obtained from microlensed events with MM $\geq$ 0.03 and  MM $\geq$ 0.1 to that from realistic population-based microlensed events in Figure~\ref{lambda_distribution}.
For the events with higher mismatch values, the ratio of the events to be discerned as microlensed ones (i.e., $\Lambda$ values located above the right bound of 2$\sigma$ C.I. of $\Lambda_{\mathcal{H}_{\rm SL}}$ in Figure~\ref{lambda_distribution}) increases.
This indicates that the residual test is valid in identifying microlensed events with high MM values.
A summary of the results is tabulated in Table~\ref{tab:result}.

\begin{figure*}[t]
    \centering
    \includegraphics[width=1.0\linewidth]{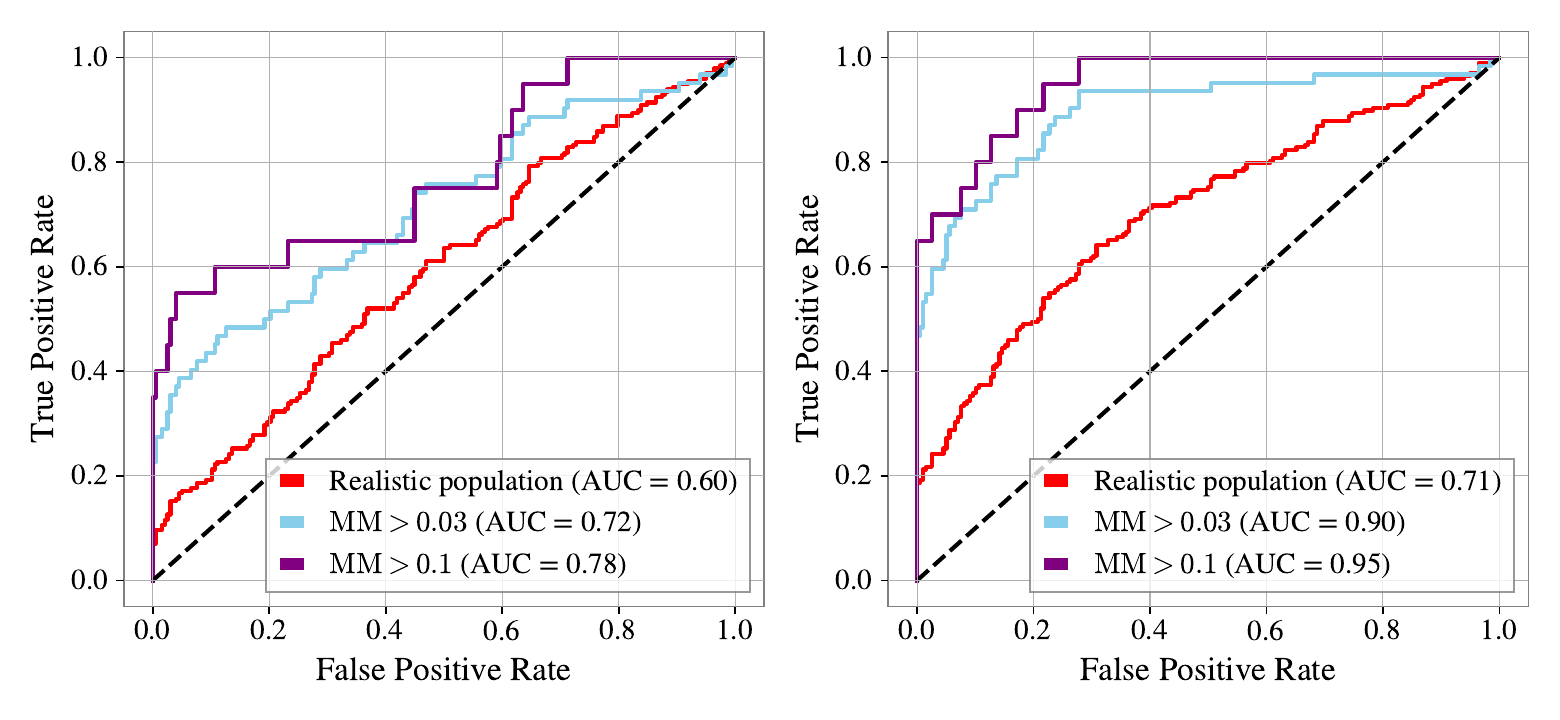}
    \caption{ROC curves for the residual test constructed by comparing each of the three event groups—$\Lambda_{\mathcal{H}_{\rm MLSL}}$ values of events drawn from a realistic population (red), events with $\mathrm{MM} \geq 0.03$ (cyan), and events with $\mathrm{MM} \geq 0.1$ (purple)—against the background distribution given by the $\Lambda_{\mathcal{H}_{\rm SL}}$ values shown in Figure~\ref{lambda_distribution}.
    The corresponding AUC values have also been estimated.
    Dashed black lines represent $\rm AUC = 0.50$, indicating random classification performance.
    The AUC values are higher for events detected with O5 sensitivity (right panels) and for those with larger mismatch values, demonstrating that microlensing detectability using residual test improves with better detector sensitivity and stronger microlensing effects.}\label{roc}
\end{figure*}

\begin{table}[t] 
    \centering
    \begin{tabularx}{1.0\linewidth}{@{}X *{4}{r} @{}}
    \toprule
        \multicolumn{4}{@{}l}{\textbf{O4 sensitivity}}\\
        \hline 
        Condition             & 
        $N_{\rm total}$ & 
        $N_{\rm ML}$ &  
        $R_{\rm ML}$ \\
        \hline
        Realistic population   & 200 & 22 & 0.110 \\
        Events with $\rm{MM}\!\geq\!0.03$ & 94 & 26 & 0.277 \\
        Events with $\rm{MM}\!\geq\!0.1$ & 35 & 12& 0.343 \\
        \hline
        \hline\\
        \hline
        \hline
        \multicolumn{4}{@{}l}{\textbf{O5 sensitivity}} \\
        \hline
        Condition & $N_{\rm total}$ & 
        $N_{\rm ML}$ &
        $R_{\rm ML}$ \\
        \hline
        Realistic population   & 200 & 43 & 0.215\\
        Events with $\rm{MM}\!\geq\!0.03$ & 94 & 49 & 0.521 \\
        Events with $\rm{MM}\!\geq\!0.1$ & 35 & 23 & 0.657 \\
    \hline
    \hline
    \end{tabularx}
    \caption{Summary of the results of residual test depicted in Figures~\ref{lambda_distribution} and~\ref{high_mismatch}.
    $N_{\rm total}$, $N_{\rm ML}$, and $R_{\rm ML}$ denote the total number of events satisfying the specified condition, the number of events discerned as a microlensed signal, and the ratio between the two numbers.}
    \label{tab:result}
\end{table}
Furthermore, we constructed receiver operating characteristic (ROC) curves to assess the overall performance of our residual-based approach.
As shown in Figure~\ref{roc}, all area under the curve (AUC) values exceed 0.5, indicating that the residual test effectively distinguishes between microlensed and unmicrolensed GW signals.
The AUC values improve with higher detector sensitivity and larger mismatch values, consistent with the trends presented in Table~\ref{tab:result2}.

\subsection{Effect of the time window on $\Lambda$}\label{sec:4.3}
In the previous sections, $\Lambda$ values were calculated using specific time windows of the form shown in Eq.~\ref{time_window}, with $\alpha=0.25$ and $\Delta t = 0.2s$, excluding the early inspiral parts of the time-domain GW waveform. 
For BBH sources, the GW signals have a lower strain in their early inspiral stage and a higher strain in the late inspiral and merger stages.
Consequently, the early parts of the time-domain data are more susceptible to noise effects compared to segments closer to the merger.   Our approach, therefore, mirrors that adopted in testing GR studies~\citep{abbott2019tests,abbott2021tests2,abbott2021tests3}, where one generally uses a narrower time window to minimize noise effects unrelated to non-GR features.
However, this approach may not be suitable in the case of microlensing signatures, as the time delays between microimages can vary. 
A narrow time window risks excluding data segments that may contain significant microlensing features, particularly in the post-merger phase.
In this context, we investigate various time windows that cover different portions of the GW signal to evaluate how noise fluctuations impact the $\Lambda$ distributions.

\begin{figure*}
    \centering
    \includegraphics[width=1.0\linewidth]{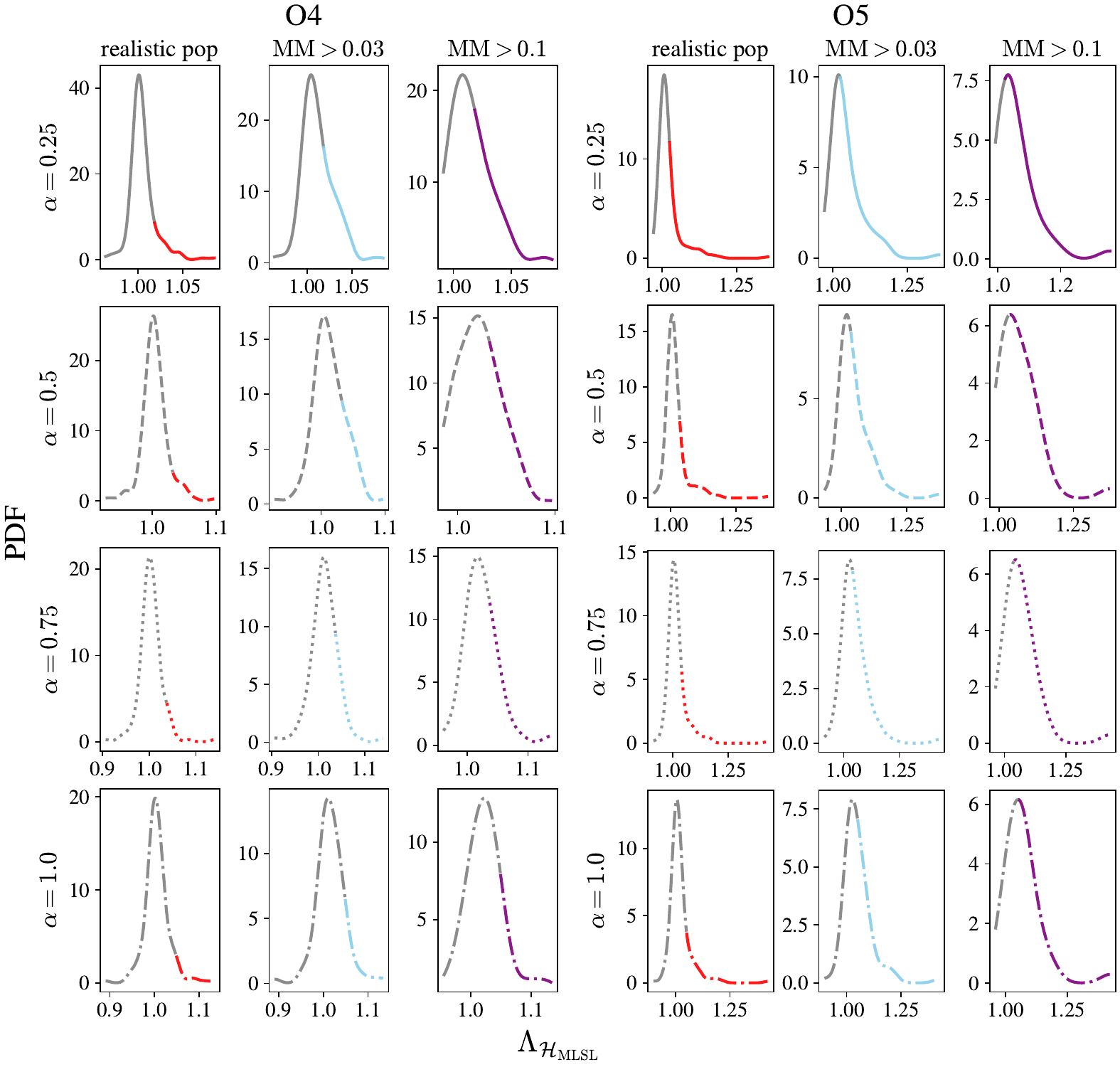}
    \caption{$\Lambda_{\mathcal{H}_{\rm MLSL}}$ distributions obtained using various time windows determined by a constant $\alpha$ and mismatch conditions (\textit{Left}: O4 sensitivity, \textit{Right}: O5 sensitivity).
    Similar to Figure~\ref{high_mismatch}, we classify microlensed events based on their mismatch values into three subsets, from left to right: realistic population-based events (red), $\rm{MM} \geq 0.03$ (green), and $\rm{MM} \geq 0.1$ (purple). 
    The distributions for $\alpha = 0.25$ (solid line) are the same as those shown in Figure~\ref{high_mismatch}, while the results for $\alpha = 0.5$ (dashed line), $\alpha = 0.75$ (dotted line), and $\alpha = 1.0$ (dash-dotted line) are obtained using time windows corresponding to half, three-quarters, and the full duration of the signals, respectively. 
    Colored segments in the distributions represent microlensed events with $\Lambda_{\mathcal{H}_{\rm MLSL}}$ falls within the range identified as microlensed, whereas grey segments indicate microlensed events misclassified as unmicrolensed. Empirically, smaller $\alpha$ values correspond to higher efficiency in identifying microlensed signals.}
    \label{fig6}
\end{figure*}
A representation of the $\Lambda_{\mathcal{H}_{\rm MLSL}}$ distributions obtained using various time windows is shown in Figure~\ref{fig6}. 
The colored segments (red, cyan, and purple) of the curves correspond to events identified as microlensed, while the gray segments correspond to events identified as unmicrolensed.
As can be seen, adopting smaller values of $\alpha$ results in the colored segments corresponding to a higher proportion of the $\Lambda_{\mathcal{H}_{\rm MLSL}}$ distributions.
This trend is, to some extent, expected because GW microlensing occurs when the time delays between repeated signals are shorter than their chirp time, and microlensing signatures are more distinctive in the higher-frequency region.
Since the merger phase and its vicinity predominantly contain high-frequency components, using a narrower time window that primarily includes the merger phase rather than the entire signal duration should enhance the efficiency of the residual test for detecting microlensing signatures.
In Table~\ref{tab:result2}, the ratios of events detected as microlensed ($R_{\rm ML}$) for each time window and mismatch condition are tabulated.
Empirically we found that using time windows with $\alpha=0.25$ yields the best efficiency for microlensing detectability.

\begin{table}[t] 
    \centering
    \begin{tabularx}{1.0\linewidth}{@{}p{3.5cm} *{4}{r} @{}}
    \toprule
        \multicolumn{5}{@{}l}{\textbf{O4 sensitivity}}\\
        \hline 
        Condition             & 
        $R_{\rm ML}$ & 
        $R_{\rm ML}^{\alpha=0.5}$&  
        $R_{\rm ML}^{\alpha=0.75}$&
        $R_{\rm ML}^{\alpha=1.0}$\\
        \hline
        Realistic population & 0.110 & 0.075 & 0.065 & 0.050\\
        Events with $\rm{MM}\!\geq\!0.03$ & 0.277 & 0.213 & 0.170 & 0.106\\
        Events with $\rm{MM}\!\geq\!0.1$& 0.343 & 0.314 & 0.257 & 0.114\\
        \hline
        \hline\\
        \hline
        \hline
        \multicolumn{5}{@{}l}{\textbf{O5 sensitivity}} \\
        \hline
        Condition & 
        $R_{\rm ML}$ & 
        $R_{\rm ML}^{\alpha=0.5}$&  
        $R_{\rm ML}^{\alpha=0.75}$&
        $R_{\rm ML}^{\alpha=1.0}$\\
        \hline
        Realistic population  & 0.215 & 0.160 & 0.155 & 0.140 \\
        Events with $\rm{MM}\!\geq\!0.03$ & 0.521 & 0.447 & 0.447 & 0.404\\
        Events with $\rm{MM}\!\geq\!0.1$ & 0.657 & 0.657 & 0.629  & 0.571\\
    \hline
    \hline
    \end{tabularx}
    \caption{The same configuration as in Table~\ref{tab:result}, but showing the ratio of events identified as microlensed to the total number of events.
    $\alpha$ denotes a constant that determines the size of the time windows.
    Narrower time windows improve the efficiency of the residual test in discerning microlensed signals.}
    \label{tab:result2}
\end{table}
\section{Discussion} \label{sec:discussion}
It is likely that strongly lensed GW signals experience microlensing effects due to stellar fields embedded in the lens galaxy, imprinting complex patterns in the GW waveforms.
Current strong lensing Bayesian inference pipelines are not designed to detect microlensing-induced distortions and consider only repeated signals with the same time-frequency evolution.
However, incorporating both strong lensing and microlensing simultaneously in joint Bayesian analysis is computationally challenging due to the increased parameter space and the complexity of calculating the amplification factors, especially in the case of stellar fields.

Our results highlight the potential of residual tests as a tool for identifying microlensing signatures in strongly lensed GW signals.
This approach can serve as a low-latency microlensing analysis pipeline, enabling us to efficiently prioritize events requiring follow-up with full microlensing analysis pipelines.
By focusing on the residuals---the difference between the recovered best-fit unmicrolensed waveform and the true microlensed signal---we can uncover GW events displaying significant microlensing signatures that would otherwise remain hidden in traditional strong lensing analyses.
The distribution of cross-correlation values for signals both strongly lensed and microlensed, compared with that for purely strongly lensed signals, suggests that microlensing effects can create statistically measurable deviations in residual tests.

As discussed by earlier studies, however, certain waveform parameters, such as spin precession and orbital eccentricity, can mimic microlensing effects by introducing similar modulations in the waveform~\citep{kim2023can,LIGO-DCC-P2300206,mishra2024exploring}, potentially leading to residuals being mistakenly attributed to microlensing.
Although we neglect these effects in this work, as scenarios involving lensing of GW signals from BBHs with strong spin precession or high eccentricity are expected to be rarer than those without such features (see e.g.,~\cite{PhysRevD.102.041302,hannam2022general,meena2022gravitational,abac2024search} for studies estimating the expected rates of those features as well as microlensing in strong lensing), distinguishing how residuals behave when the modulations arise from microlensing, orbital eccentricity, or spin precession should be addressed in future work.

In the context of our cross-correlation calculation, we utilized maximum likelihood waveforms to obtain a single statistic per event. However, a more accurate approach would involve using waveforms corresponding to all posterior values obtained, which would thus generate a distribution for the cross-correlation value.
One could then assess the degree of overlap between this distribution and the full cross-correlation distribution obtained from all events.
Furthermore, Gaussian noise realizations were generated to establish the baseline for the distribution of cross-correlation values, which is valid for general cases.
Nevertheless, in actual detection cases, detector noise may involve non-Gaussian features, such as glitches.
Therefore, processes such as deglitching, which aim to mitigate non-Gaussianity as much as possible, must be performed prior to conducting the residual test, though if microlensing is not accounted for, some features may be removed in this procedure.

When considering the entire dataset containing microlensed events with both low and high mismatch, the results for aLIGO O4 sensitivity suggest that it may not be sufficient to reliably detect microlensing effects. 
However, as shown by the results using aLIGO O5 sensitivity, the improved sensitivity can be expected to enhance our ability to discern microlensing signatures in signals that are both strongly lensed and microlensed.

It is also important to note that, while we employed one of the state-of-the-art microlensing models and corresponding parameter ranges to simulate microlensing fields embedded in a galaxy as a baseline, other representative microlensing population models and macromodel magnifications could also be used. 
In general, a higher macromodel magnification and a higher density of microlenses would result in larger mismatches and therefore, a higher chance of detection. 
For investigations into how the macromodel plays a role in microlensing across parameter spaces, we refer the reader to~\cite{diego2019observational,mishra2021gravitational,Shan:2022xfx}. 
However, in addition to the baseline estimates considering a realistic population, we also quantify our ability to detect microlensing by looking at selected high-mismatch waveforms (Table~\ref{tab:result}), so that our results can also be interpreted in the context of these mismatches. 
We leave the full investigation of various microlensing fields in conjunction with galaxy lens models as future work.

When considering only the high-mismatch events that the current microlensing search pipeline can detect, the ability of the residual test to identify microlensed signals is found to improve for both O4 and O5 sensitivity.
However, there is a caveat: excessively distorted microlensed signals may experience significant SNR loss during the matched-filtering process if microlensing effects on the templates are not properly accounted for -- see \cite{chan2024detectability} for details.
Therefore, a dedicated matched-filtering search pipeline incorporating microlensed templates may prove to be required.

In the era of next-generation detectors, with a higher detection rate of lensed GW signals, such low-latency methods will become more important.
We anticipate our approach to be more efficient for next-generation detectors due to longer-duration signals, louder SNRs and improved sensitivity at high frequencies.
A detailed study to quantify the efficiency in these scenarios will be carried out in future work.

A single joint parameter estimation followed by the residual test for a pair of lensed events takes approximately half a day, given that single likelihood evaluation typically takes only a few milliseconds in our approach. 
In contrast, when using a frequency-domain source model that incorporates both strong lensing and microlensing effects in a typical GW search pipeline, each likelihood evaluation for our simulated waveform within a stochastic sampler such as nested sampling takes approximately 10 seconds, roughly $10^{4}$ times slower.
The residual test presented in this work has the potential to serve as a fast and low-latency pre-analysis tool, alerting us to possible microlensed events before more computationally intensive full analyses are conducted.

\section*{acknowledgments}
We thank Anupreeta More and Alvin K.Y. Li for their valuable comments and feedback.
E. S. is supported by grants from the College of Science and Engineering of the University of Glasgow.
X. S. is supported by the National Natural Science Foundation of China (Grant Nos. U1931210, 11673065, and 11273061).
O.~A.~H. acknowledge support by grants from the Research Grants Council of Hong Kong (Project No. CUHK 14304622 and 14307923), the start-up grant from the Chinese University of Hong Kong, and the Direct Grant for Research from the Research Committee of The Chinese University of Hong Kong. 
M. H. is supported by the Science and Technology Facilities Council (Grant Ref. ST/V005634/1).
B. H. is supported by the National Key R\&D Program of China No. 2021YFC2203001.
The authors are grateful for computational resources provided by the LIGO Laboratory and supported by the National Science Foundation Grants PHY-0757058 and PHY-0823459.
This material is based upon work supported by NSF's LIGO Laboratory which is a major facility fully funded by the National Science Foundation
\newpage
\bibliography{residual}{}
\bibliographystyle{aasjournal}

\appendix
\section{Mock data simulation}\label{app:mock_data}
For our mock dataset, we first simulate binary black hole (BBH) events using a Monte Carlo method based on a model for the theoretical merger rate evolution which is proportional to the star formation rate (SFR), as parametrized in \cite{madau2014cosmic} with a 50 Myr delay between star formation and BBH formation (see Appendix B of ~\citet{Xu:2021bfn} for more details).
We adopt a \textsc{PowerLaw + Peak} distribution with population hyperparameters and parameter ranges given in~\citet{LIGOScientific:2018jsj} to sample the primary and secondary masses ($m_{1}, m_{2}$) and assume an inclination angle distribution of $p(\iota) \propto \sin(\iota)$.
Other parameters, including component spins ($\vec{a}_1, \vec{a}_2$), polarization angle ($\psi$), right ascension ($\alpha$), declination ($\delta$), coalescence time ($t_c$), and coalescence phase ($\phi_c$), are drawn from uniform distributions within their respective prior ranges.

Next, we select a subset of BBH events to be classified as strongly lensed. 
Specifically, we calculate the optical depth of multiple images for each BBH event assuming a singular isothermal sphere (SIS) model~\citep{Haris:2018vmn}\footnote{We adopt the SIS optical depth for simplicity, acknowledging an approximate 5–10\% difference~\citep{wierda2021beyond}, as our objective is not to derive a precise estimate of the number of lensed events.} for our macrolens.
A uniformly distributed random number between 0 and 1 is assigned to each event, and if the calculated optical depth $\tau$ exceeds this value, the event is categorized as a strongly lensed GW signal; otherwise, it is excluded.
For each selected event (i.e., $h_{\rm SL}$), strong lensing parameters are assigned following the singular isothermal ellipsoid (SIE) model, which includes the axis ratio ($q$), velocity dispersion ($\sigma_v$), and impact parameter ($y$).
The velocity dispersions of the lensing galaxies are sampled from SDSS galaxy distributions~\citep{2015ApJ...811...20C} and the corresponding axis ratio values are sampled from a Rayleigh distribution with the corrected scaling parameter from~\citet{wierda2021beyond}.
The impact parameters are drawn from a uniform distribution.
Note that some lensing systems produce quadruple images; in such cases, we select the first two images that arrive to ensure that each lensing system results in two detectable GW signals.

Third, to generate both strongly lensed and microlensed GW signals (i.e. $h_{\rm MLSL}$), we simulate microlensing fields assuming a stellar mass distribution that follows the late Salpeter initial mass function (IMF)~\citep{spera2015mass} and a Sérsic radial profile~\citep{vernardos2019microlensing}. 
For consistency with~\citet{Diego:2021mhf,shan2023microlensing}, we set the IMF to include remnant objects, assuming that their mass density constitutes 10\% of the total stellar mass density and microlens masses are sampled within the range $0.1 - 1.5$ solar masses.

Finally, we construct a dataset containing 100 pairs of strongly lensed GW signals ($h_{\rm SL}$) and 100 pairs of signals that are both strongly lensed and microlensed ($h_{\rm MLSL}$).
Additionally, we increase the magnification factors of the macrolens to $\mu = 100$ in order to obtain signals which are significantly microlensed, and which therefore do not follow our realistic population.
This is done to validate our approach (see Sec.~\ref{sec:4.2}).
Events are considered detectable if their network signal-to-noise ratio (SNR) exceeds 12. 
For this analysis, we assume a detector network consisting of the two LIGO observatories in Hanford and Livingston, as well as the Virgo detector in Pisa, operating at O4 sensitivity.
\end{document}